\documentclass[english,twocolumn,prb,preprintnumbers,amsmath,amssymb,superscriptaddress,nofootinbib]{revtex4}

\usepackage{verbatim}
\usepackage{graphicx}
\usepackage{amssymb}
\usepackage{babel}

\makeatother

\newcommand{\RN}[1]{%
	\textup{\uppercase\expandafter{\romannumeral#1}}
}
\begin{document}
	
	%\preprint{XXX}

	\title{Motion Transduction with Thermo-mechanically Squeezed Graphene Resonator Modes}
	
	\author{Rajan Singh}
	\affiliation{Department of Physics, Indian Institute of Technology - Kanpur, UP-208016, India}
	
	\author{Ryan J.T. Nicholl}
	\affiliation{Department of Physics and Astronomy, Vanderbilt University, Nashville, Tennessee 37235, USA}

	\author{Kirill Bolotin}
	\affiliation{Department of Physics, Freie Universitat Berlin, Arnimallee 14, Berlin 14195, Germany}
	
	\author{Saikat Ghosh}
	\email{gsaikat@iitk.ac.in}
	\affiliation{Department of Physics, Indian Institute of Technology - Kanpur, UP-208016, India}

%	
%	\date{\today}

	%\pacs{xxx}

	
	\begin{abstract}
		
	\noindent
		\textbf{ABSTRACT}: There is a recent surge of interest in amplification and detection of tiny motion in the growing field of opto and electro mechanics. Here, we demonstrate widely tunable, broad bandwidth and high gain all-mechanical motion amplifiers based on graphene/Silicon Nitride (SiNx) hybrids. In these devices, a tiny motion of a large-area SiNx membrane is transduced to a much larger motion in a graphene drum resonator coupled to SiNx. Furthermore, the thermal noise of graphene is reduced (squeezed)  through parametric tension modulation. The parameters of the amplifier are measured by photothermally actuating SiNx and interferometrically detecting graphene displacement. We obtain displacement power gain of 38 $\rm dB$ and demonstrate 4.7 dB of squeezing resulting in a detection sensitivity of 3.8 $\rm fm/\sqrt{Hz}$, close to the thermal noise limit of SiNx. \\
		\noindent
		\textbf{KEYWORDS}: \textit{Graphene, Nems, Transducer, Parametric Amplification, Thermomechanical Squeezing.}
	\end{abstract}

	\maketitle	
	
Measuring small forces with high precision has been a major goal in fields ranging from gravitational astronomy to atomic force microscopy~\cite{Braginsky92,Abbott16,Binnig86}. Mechanical resonators of low mass and high quality factor ($Q$), that operate by converting force into displacement, have been a dominant choice for such detectors. Resonators based on two-dimensional (2D) materials such as graphene~\cite{Bunch07,Poot12}, with $Q>2\times10^4$ at cryogenic temperatures, have enabled detection of forces at a record level, down to zeptonewtons~\cite{Weber16,Singh14}. However, $Q$ of graphene resonators decreases by orders of magnitude at room temperatures. Alternatively for room temperature operation, more traditional Silicon Nitride (SiNx) resonators with $Q$ above $2\times 10^7$ and force sensitivity of attonewtons have been developed~\cite{Norte16,Reinhardt16}. Significantly larger mass of these resonators compared to graphene leads to small amplitude of the force-induced motion. Accordingly, a variety of transducers have been proposed to amplify tiny motion~\cite{Caves82,Clerk04,Clerk10}. Traditionally, optical interferometers have enabled detection of resonator displacements at unprecedented precision. At the same time, high  optical power used in such interferometers often leads to heating and is particularly severe for micro-resonators in restricted on-chip environment. A variety of  auxiliary on-chip motion pre-amplifiers and transducers have been proposed and demonstrated to improve detection sensitivity without additional heating. These include tunnelling point contacts~\cite{Knobel03,LaHaye04,Poggio08}, superconductors with capacitive coupling~\cite{Singh14,Weber16,Etaki08} and microwave cavities~\cite{Regal08}, all demonstrating high gain and sensitivity~\cite{Clerk04,Clerk10}. However, reliable and robust amplification of motion remains challenging. In particular, stringent coupling requirements of these transducers to the target resonator result in low device yield. Furthermore resonant nature of such amplifiers designed at a specific frequency compromises on tunability, and their high $Q$ leads to a narrow detection bandwidth.
	
	Here we realize a highly tunable, broad bandwidth, high-gain, all-mechanical motion amplifier~\cite{Mahboob10,Huang13} at room temperature that is based on SiNx/graphene hybrids. Our goal is to amplify motion of the target resonator, a large area SiNx membrane. A monolayer of graphene is deposited onto holes etched on the SiNx. The supported part of graphene is coupled to the SiNx by robust Van der Waals forces. The graphene pre-amplifier operates as follows: by electrostatically controlling the tension of the suspended graphene resonator, its mechanical modes are brought into resonance with the target mechanical mode of SiNx. Coupling between the resonators together with large disparity of their masses leads to amplification: a tiny motion of SiNx is converted into much larger motion of graphene, which is eventually detected by optical interferometry. Comparing SiNx to graphene displacement power spectra, we measure amplification with an average power gain of 36.8 dB in 5 resonators. Furthermore, by modulating  the intrinsic tension of graphene, we demonstrate parametric gain accompanied by $4.7$ dB suppression (squeezing) of graphene's thermal noise in one motion quadrature. With additional feedback squeezing of the thermal noise~\cite{Vinante13} along with side-band cooling of the amplifier mode temperature~\cite{Alba16,Mathew16}, graphene-based motion amplifiers can reach unprecedented displacement sensitivities at room temperature.
	\begin{figure*} 
		\includegraphics[scale=0.80]{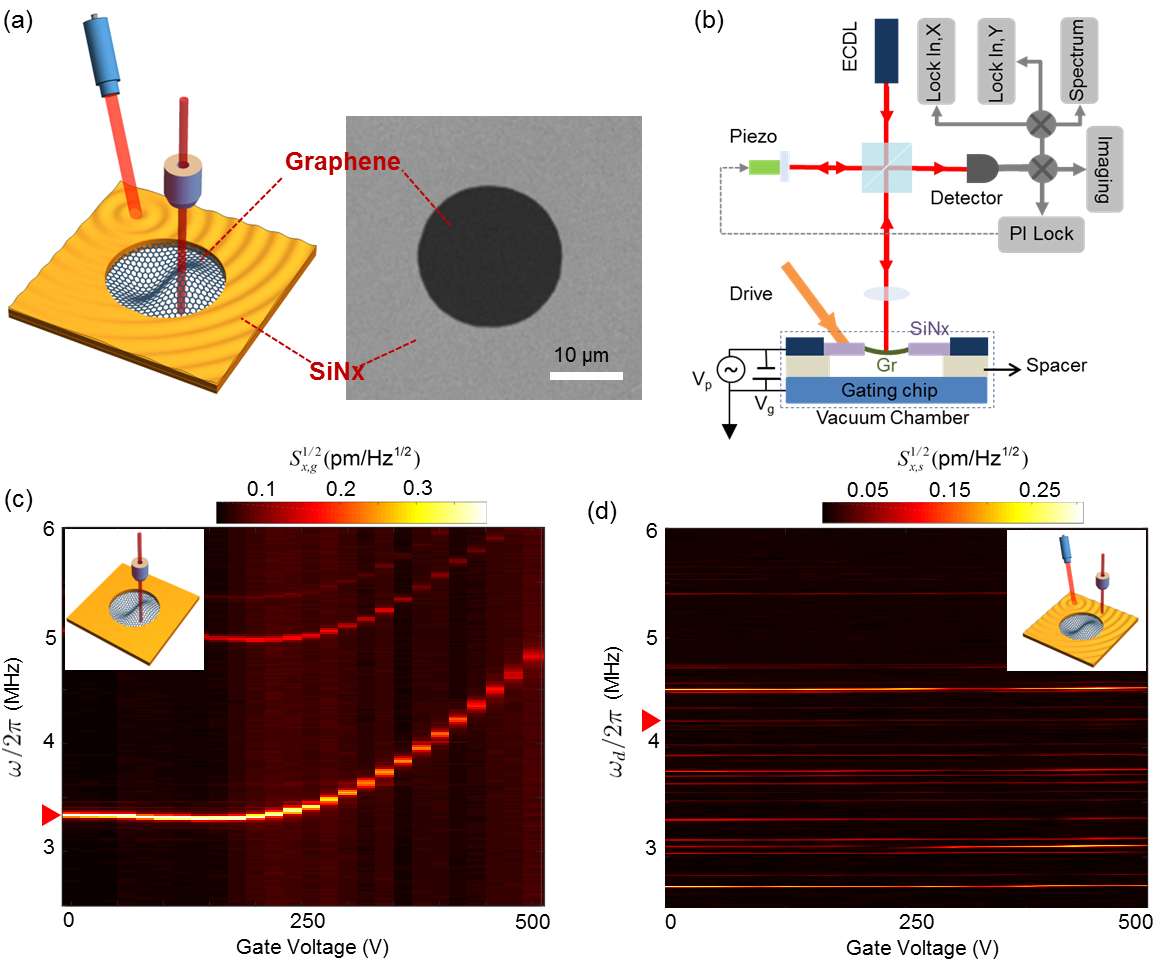}
		\caption{ \textbf{Graphene and SiNx resonators:}
			\textbf{(a)} Scanning electron microscope image of a typical device together with its cartoon representation. \textbf{(b)} Schematic of the experimental setup. A frequency and power stabilized external cavity diode laser acts as a probe for a confocal microscope in an interferometric configuration. One arm of the microscope is actively stabilized. Graphene is actuated by applying a gate voltage between it and the substrate whereas SiNx is actuated by periodically heating it with an external laser beam. Light reflected from graphene or SiNx is detected by a photodetectors and analyzed either by a spectrum analyzer or a lock-in amplifier. \textbf{(c)} Brownian displacement  spectral density $S_{x,g}^{1/2}$ of an undriven $20$ $\rm \mu m$ diameter graphene resonator (device A), as a function of probing frequency and gate voltage(measured at a probe power of 0.4 mW). \textbf{(d)} Displacement  spectral density $S_{x,s}^{1/2}$ for a SiNx membrane driven by an external laser beam. Red arrows indicate interacting modes of graphene and SiNx resonators (probe power of 0.4 mW).}
	\end{figure*}

	\begin{figure*}\includegraphics[scale=0.8]{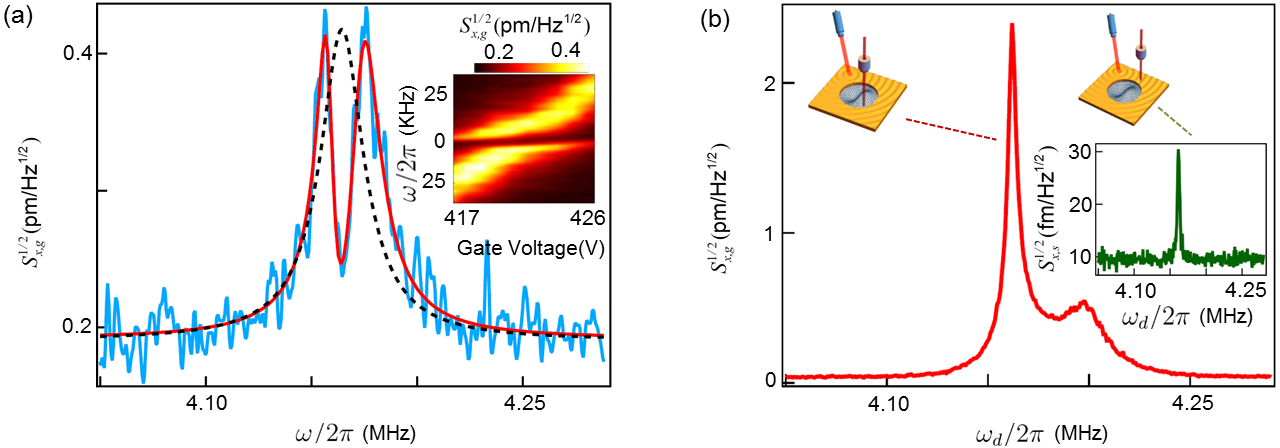}
		\caption{\textbf{ Graphene/SiNx coupling and gain:} 
			\textbf{(a)} Blue curve is a displacement spectrum of undriven graphene at $ V_g = 420$ $\rm V$, close to resonance with SiNx (probe power of 0.1 mW). Red curve is a fit to the data using the bi-linear interaction model described in the text. \textbf{(inset)} Corresponding avoided level crossing with static gate voltage (detuning).   \textbf{(b)} Displacement spectrum measured on graphene (red curve) and on SiNx (green curve and separately in the inset) for driven SiNx (probe power of 0.4 mW). A gain of $G_{c} \sim 38$ $\rm dB$ is extracted by comparing the two spectra. The broad shape of the peaks in the red curve is set by thermal spectrum of graphene, dominating the total added noise. It can be noted that the noise floor for SiNx (inset) is lower than that of graphene due to difference in reflectivity of the two materials~\cite{SI}.} 
		
	\end{figure*}
	
%\noindent
\emph{Experimental setup:} At the heart of the experiment is
a large area (320$\times$320 $\times$0.3 $\mu$m$^{\rm3}$) SiNx membrane covered by a 20 nm-thick layer of gold (Au), with through holes of diameters 15 and 20 $\mu$m etched in it. A monolayer CVD graphene is mechanically transferred onto the entire structure. A conductive silicon chip is placed 30 $\rm \mu m$ below graphene/SiNx and separated from it by a layer of insulating material \cite{Nicholl09}. Separate electrical contacts are made to SiNx/Au/graphene and bottom Si and the entire structure is placed inside a vacuum chamber. All measurements are done at room temperature and $10^{-2}$ $\rm mbar$ pressure. While measurements were done on five devices, we focus on the data from two of them. The built-in tension of SiNx in device A is 600 MPa, estimated from rather sparse vibrational modes of that device (Fig. 1d). Device B has lower tension of 150 MPa, with denser distribution of vibrational modes~\cite{SI}.
	
	Mechanical motion of suspended graphene and SiNx membranes is independently actuated and detected. To probe the motion, a fiber-based confocal microscope is focused on either graphene or SiNx, while weak motion-induced modulation in the reflected signal is amplified and recorded by an interferometer in Michelson configuration (Fig. 1b).  Graphene is actuated electrostatically by applying an oscillating voltage $V_g$ between it and the gating chip. This modulation is too small to drive heavier SiNx  membrane. Instead, SiNx is driven photothermally by periodically heating it with laser beam of optical power $\sim 3.4$ mW incident at an angle.
	
	\emph {Graphene and SiNx resonances:} When the microscope is focused on graphene, distinct peaks are visible in the power spectrum of the reflected light, even when the membrane is not driven externally (Fig. 1c). These peaks correspond to the resonant drum modes of suspended graphene excited by thermal fluctuations~\cite{Davidovikj16}. We use the equipartition theorem to convert the measured optical power spectrum into displacement power spectrum~\cite{Hauer13,SI}. For the fundamental mode, the root mean square displacement is around 30 pm. Along with the fundamental mode ($\omega_0/2\pi=3.32$ MHz at $ V_g = 0$) several higher order modes are typically visible~\cite{SI}. The modes match well to simulated modes of circular membrane with anisotropic tension and an average quality factor $Q_g \sim$ 200. When a static gate voltage $V_g$ is applied, the mode frequencies upshift, indicating increased tension~\cite{Chen09} (Fig. 1c). We fit the corresponding dispersion curve to extract the effective mass ($m_g=29\times m_0$, where $m_0 $ is the effective mass of single layer graphene resonator) for the fundamental mode.~\cite{Chen09, Chen13}
	
	For the significantly heavier SiNx membrane, the amplitude of thermal motion is below our interferometer detection limit. To convert the detected power spectrum on SiNx to displacement, we use the thermal spectra of graphene correcting for difference in surface reflectivity while carefully maintaining all other detection parameters. When the membrane is driven photo-thermally, multiple peaks with small frequency spacing and higher quality factors ($Q_s$ $\sim$ 3,000) are visible (Fig. 1d). The frequencies of these modes are nearly independent of $V_g$, as the tension induced by electrostatic pulling is significantly smaller compared to the built-in tension of SiNx. Simulation suggests that these modes correspond to  modes of a square membrane~\cite{SI}.

\begin{figure*}\includegraphics[scale=0.8]{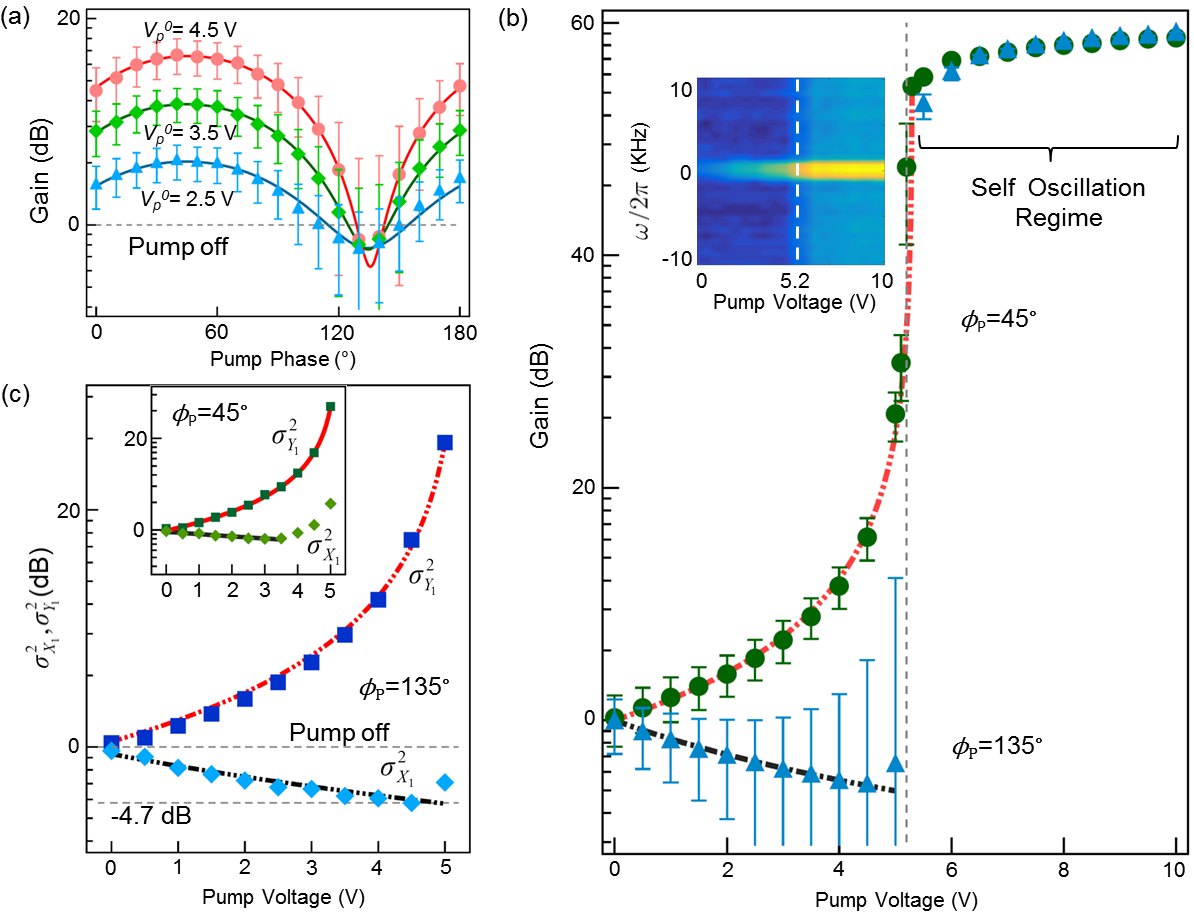}
	\caption{\textbf{Parametric squeezing: (a)} Parametric gain vs. phase of the pump signal (with respect to a weakly driven signal) for device B modulated (pumped) at twice its resonant frequency for pump amplitudes $V_p^0=2.5$ V, $3.5$ V, and $4.5$ V.  \textbf{(b)}  Amplification (at $\phi_p=45^\circ$) and deamplification (at $\phi_p=135^\circ$) of a signal vs. pump voltage. Self-oscillation sets in at a critical voltage $V_{p}^c=5.2$ $\rm V$ (inset). These curve are fitted~\cite{SI} with
		function $G_p{(\phi_p=45^\circ,135^\circ)}=1/(1\mp\frac{V_p^0}{V_{p}^c})$. \textbf{(c)} Normalized variance in graphene quadratures  $X_1$ and $Y_1$  vs. pump voltage at $\phi_p=135^\circ$. Similar plot, but for $\phi_p=45^\circ$ is shown in the inset. Signal goes into a self-oscillation regime above $V_{p}^c$ with overall increase in noise floor.}   
\end{figure*}

	We use electrical gating to bring one mode of graphene in resonance with a mode of SiNx. Such resonance conditions are fulfilled for modes marked by red triangles in Figs. 1c,d at $\omega/2\pi= 4.17$ MHz. 
	Around the degeneracy, the modes are coupled and hybridized~\cite{Schwarz16,Okamoto13}. Indeed, we observe a sharp splitting in the Brownian mode of graphene (Fig. 2a, blue curve).  Moreover, two split peaks exhibit an avoided-crossing pattern vs. $V_g$ around degeneracy, a tell-tale sign of inter-mode interaction (Fig. 2a, inset).

 \emph{All-mechanical amplifier:} We now focus on the use of graphene as an all-mechanical amplifier~\cite{Kim09} of SiNx motion. Experimentally, we observe amplification in device A by weakly driving the SiNx mode photo-thermally, while recording the power spectrum both on SiNx and graphene. When graphene mode is not in resonance with SiNx, photo-thermal driving of SiNx {\it does not} affect the spectrum of graphene. However, on resonance, we observe a peak in graphene displacement spectrum $S_{x,g}^{1/2}$ (Fig. 2b, red curve) which is much larger than that measured on SiNx, $S_{x,s}^{1/2}$ (Fig. 2b, green curve and inset). Ratio of the two peak heights is the coupling power gain of our all-mechanical amplifier $G_{\rm c}=S_{x,g}/S_{x,s} = 6.2\times 10^3$ (38 dB).  We find an average gain of $36.8$ dB across the range of 5 measured devices. We note that the frequency-dependent gain spreads over a relatively large band-width of $\sim$ 38 kHz(FWHM)~\cite{SI}.

To quantitatively analyze the amplifier parameters, we use a model of graphene and SiNx resonators (transverse displacements $x_g$ and $x_s$) interacting via bilinear coupling $\alpha x_g x_s$~\cite{SI, Okamoto13}. The Brownian noise spectrum of graphene/SiNx  fits  well with the model (red trace in Fig. 2a) with $\alpha/4\pi^2$ $\sim$ $11.9(\pm0.2)\times 10^{-3}~{\rm kg~Hz}^2$ for device A. Furthermore from fitting of the driven SiNx mode (Fig.2b), the extracted gain is $G^{\rm fit}_{\rm c} =$ 34.6 dB.
  
An approximate, on-resonant expression for the power gain of $G_{\rm c}^{\rm th} \sim \left|\frac{\alpha}{m_g\gamma_g\omega_0}\right|^2$ (42 dB) scales directly with coupling constant, inversely with effective mode mass $m_g$ and damping $\gamma_g$ of graphene and is in agreement with the measured value. 
It can be noted that  $G_{\rm c}^{\rm th} = Q_g^2$ ($\sim$ 46 dB) sets the maximum limit of the power gain~\cite{Schwarz16,SI}. 
	
\begin{figure*}\includegraphics[scale=0.8]{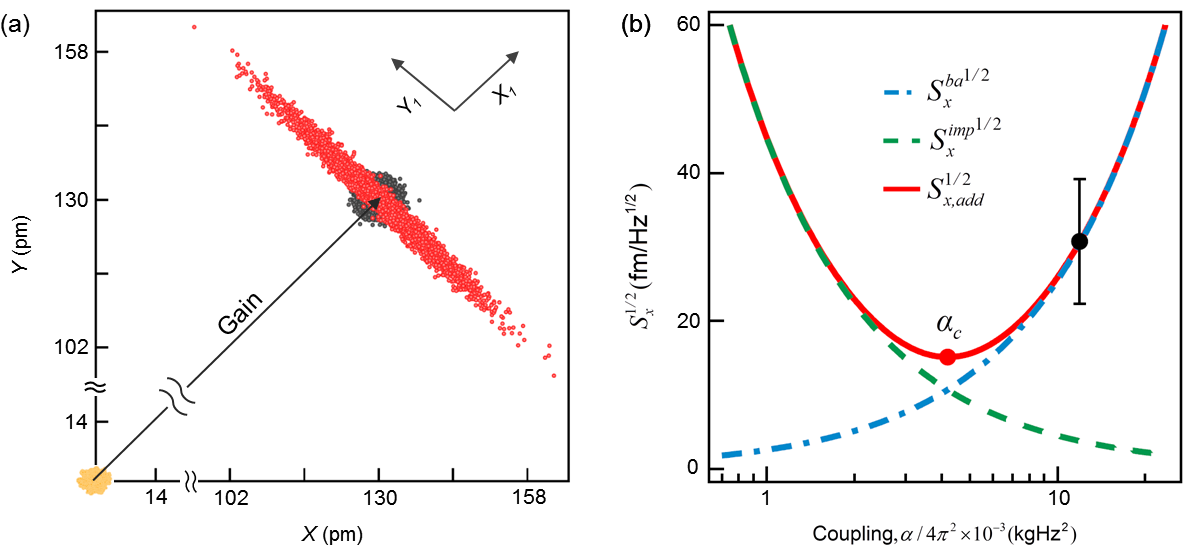}
	\caption{\textbf{Noise in graphene pre-amplifier:(a)} Each point is an independent measurement of $X$ and $Y$ quadratures. The points cloud correspond to detector focused on weakly driven SiNx (yellow), on graphene (grey) and on parametrically driven graphene (red). The cloud points are experimental data, while the cloud centers are representative values corresponding to measured amplifier parameters. \textbf{(b)} Total added noise (red) composed of detector imprecision noise (green dash) and back action (blue dash), as a function of  coupling constant. One data point with error bars illustrates typical parameters for the measured devices.}   
\end{figure*}
	
	Detection sensitivity of the graphene amplifier is limited by the noise of interferometric detector $S_{I_0}^{1/2} \sim 0.04 \rm - 0.20$ $\rm pm/\sqrt{Hz}$ (for probe power of 0.4 mW to 0.1 mW, incident on graphene) along with the Brownian noise added by graphene, ${S_{x,g}^{\rm th}}^{1/2} \sim$ 0.38 $\rm pm/\sqrt{Hz}$. When referred back to minimum detectable displacement on SiNx, the resultant sensitivity (imprecision noise) is
	\begin{equation}
		{S^{\rm imp}_{x}}^{1/2} =\Big(\frac{ S_{I_0}(\omega)+ S^{\rm th}_{x,g}}{G_{c}}\Big)^{1/2}.   \\ \nonumber 
\end{equation}

For device A, we estimate ${S^{\rm imp}_x}^{1/2}\sim 5.5(\pm0.7)$ $\rm fm/\sqrt{Hz}$. It is dominated by the Brownian motion of graphene. Further improvement therefore requires either a boost in gain or a reduction in thermal fluctuations of graphene.  
\\

\emph{Squeezing of thermo-mechanical noise:} We add a second stage of amplification by parametrically driving graphene resonators~\cite{Rugar91,Prasad17,EichlerN11,Natrajan95}. Parametric amplification is studied in device B by actuating SiNx with a weakly driven resonant signal. Tension of the device is modulated at twice its resonant frequency by applying an AC pump voltage $V_p = V_p^0\cos(2\omega_0 t+\phi_p)$ of amplitude $V_p^0$ and phase $\phi_p$ ($\phi_p=0$ corresponds to the phase of the weakly driven signal on SiNx) between graphene and the back gate. Comparing the displacement power spectra of graphene with ($S^{V_p}_{x,g}$) and without ($S^{V_p=0}_{x,g}$) pump signals, we observe an additional {\it parametric} gain in power spectrum measured on graphene $ G_{p}(V_p,\phi_p)=S^{V_p}_{x,g}/S^{V_p=0}_{x,g} $ (Fig. 3a). This gain is phase($\phi_p$)-dependent, with  maximum at $\phi_p =45^\circ$ and minimum at $\phi_p = 135^\circ$. With increasing pump $V_p^0$, the graphene enters into a self-oscillation regime at a critical voltage  $V^{c}_p\sim$ 5.2 V (Fig. 3b and inset). 

In presence of the pump, fluctuations in the signal (graphene displacement) become phase-dependent as well. This is particularly evident when we measure the two quadratures $X$ and $Y$ of graphene motion, defined as $x_g(t) = X\cos(\omega_0 t)+Y\sin(\omega_0 t)$, with a dual channel lock-in amplifier. The pump signal is in phase with $X$ for $\phi_p=0$. With increasing pump voltage $V_p^0$, fluctuations in the quadratures become correlated. Specifically, the fluctuations are minimum for the combination $X_1=1/\sqrt2(X+Y)$ and maximum for $Y_1=1/\sqrt{2}(Y-X)
$. Variance $\sigma_{X_1}^2$ and $\sigma_{Y_1}^2$ for $X_1$ and $Y_1$ are plotted in Fig. 3c (normalized with respect to the variance $\sigma_0^2$ without pump) for pump parameters $\phi_p$ = 135$^\circ$ and $V_p^0 = 4.5$ $\rm V$. We observe
 a decrease of 4.7 dB in $\sigma_{X_1}^2$ when compared to its value in absence of pump. This is a clear signature of parametric squeezing. To the best of our knowledge, this is the first observation of thermo-mechanical squeezed states in 2D resonators. 
 
The observed squeezing improves performance of the amplifier. The corresponding detection sensitivity lowers due to reduction of the thermal fluctuations along $X_1$. Using squeezing parameters in Fig. 3c at $V_p = 4.5$ $\rm V$ as typical values, the effective gain is $\mathcal{G}=G_{p}\times G_{\rm c}\sim$ 32.3 dB  and the estimated measurement sensitivity is $3.8(\pm0.5)$ $\rm fm/\sqrt{Hz}$. Instead of optimizing sensitivity, one can also simply boost up the overall gain to  $\sim$ 47 dB (for parameters in Fig. 3c, inset at $V_p = 3.5$ $\rm V$) with a corresponding large modulation to detect. However, in this case the squeezing is less, with the sensitivity degrading to $4.6(\pm0.6)$ $\rm fm/\sqrt{Hz}$.\\

{\emph{Discussion:}} Performance of the proposed graphene pre-amplifier along with relative contributions of various noise sources is summarized in Fig. 4a. Each point represents a pair of quadrature values $X$ and $Y$, integrated over 20 ms. The yellow cluster correspond to displacements detected on SiNx, where the spread is set by the detector noise (larger than the SiNx thermal motion). The red cloud correspond to displacements measured on parametrically driven graphene (For device B, with $V_p^0 = 4.5$ $\rm V$, $\phi_p = 135^\circ$). The elliptical shape of the data cluster indicates squeezing. Squeezing is especially evident in comparison with the circular black cloud  due to fluctuations of data point measured on  un-pumped graphene. To show gain, the center of the red cluster is shifted with respect to yellow.

Relative strengths of intrinsic noise~\cite{Clerk10} contributions of the graphene pre-amplifier, on SiNx, are summarized in Fig. 4b. First, the thermal fluctuations of graphene produce back-action force on the target resonator (SiNx). The corresponding contribution in displacement spectral density, ${S^{\rm ba}_x}^{1/2}= (\frac{\alpha}{m_s\gamma_s\omega_s}) {S^{\rm th}_{x,g}}^{1/2}$ scales directly with coupling strength $\alpha/4\pi^2$ and inversely with the mass of graphene (Fig. 4b, blue curve). On the contrary, the imprecision noise (${S^{\rm imp}_{x}}^{1/2}$), scales down with $\alpha/4\pi^2$ but increases with $m_g$. Therefore, there exists a critical coupling strength ($\alpha_c/4\pi^2$) or, equivalently, a critical mass (number of graphene layers $n_c$) for which the amplifier is quietest (adds minimum noise). For typical values of our devices (black point in Fig. 4b), the parameters are close to optimal, with thermal back action force of graphene dominating the noise on SiNx. 

Our results suggest several routes to further improve the performance of graphene pre-amplifiers. First, by controlling the mass or built-in tension, the noise sources of the resonators can be moved closer to the optimal intersection point in Fig. 4b. With further squeezing using electronic feedback~\cite{Vinante13}, one can go beyond the 6 dB (thermal) limit, thereby improving sensitivity without adding any back-action noise~\cite{Clerk10}.

In conclusion, we have demonstrated that graphene can serve as an all-mechanical amplifier for a coupled mode of a SiNx resonator. The amplifier has a gain of 38 dB, bandwidth of 38 kHz and an overall detection sensitivity of 3.8 $\rm fm/\sqrt{\rm Hz}$, limited by graphene thermal fluctuations. With further squeezing of thermal noise through feedback, along with choice of optimal device parameters, graphene based motion amplifiers can serve as an important tool in the growing field of opto and electro-mechanics.

\noindent\\
\textbf{AUTHOR INFORMATION:}\\
\textbf{Corresponding Author} \\   
*Email: gsaikat@iitk.ac.in\\
\textbf{Notes}\\
The authors declare no competing financial interest.

\noindent\\
\textbf{ACKNOWLEDGMENTS:}\\
%\section*{Acknowledgment:}
We thank C. S. Vatsan, Sagar Chakraborty, Amit Agarwal, C. S. Sundar, H. Ulbricht, Mishkatul Bhattacharya and Jan Kirchoff for insightful discussions and comments. We also thank Om Prakash for his numerous help in construction of the experimental setup. This work was
supported under DST Grant No. SERB/PHY/2015404 and ERC Grant No. 639739.\\\\
\\

%\textbf{AUTHOR INFORMATION:}\\
%\textbf{Corresponding Author} \\   
%*Email: gsaikat@iitk.ac.in\\
%\textbf{Notes}\\
%The authors declare no competing financial interest

%% LyX 1.4.3-5 created this file.  For more info, see http://www.lyx.org/.
%% Do not edit unless you really know what you are doing.
%\documentclass[english,onecolumn,prb,preprintnumbers,amsmath,amssymb,superscriptaddress]{revtex4}
%%\documentclass[english,preprint,preprintnumbers,amsmath,amssymb,superscriptaddress]{revtex4}
%%\usepackage[T1]{fontenc}
%%\usepackage[latin1]{inputenc}
%\usepackage{verbatim}
%\usepackage{graphicx}
%\usepackage{amssymb}
%\usepackage{babel}
%\usepackage{titlesec}
%%\usepackage{xcolor}
%\makeatother
%
%\setcounter{equation}{0}
%\renewcommand{\theequation}{S.\arabic{equation}}
%\renewcommand\thefigure{S.\arabic{figure}}    
%\setcounter{figure}{0} 

\setcounter{equation}{0}
\renewcommand{\theequation}{S.\arabic{equation}}
\renewcommand\thefigure{S.\arabic{figure}}    
\setcounter{figure}{0} 

%\makeatletter

%%%%%%%%%%%%%%%%%%%%%%%%%%%%%% LyX specific LaTeX commands.
%% Bold symbol macro for standard LaTeX users
%\providecommand{\boldsymbol}[1]{\mbox{\boldmath $#1$}}

%%%%%%%%%%%%%%%%%%%%%%%%%%%%%% User specified LaTeX commands.
% ****** Start of file apssamp.tex ******
%
%   This file is part of the APS files in the REVTeX 4 distribution.
%   Version 4.0 of REVTeX, August 2001
%
%   Copyright (c) 2001 The American Physical Society.
%
%   See the REVTeX 4 README file for restrictions and more information.
%
% TeX'ing this file requires that you have AMS-LaTeX 2.0 installed
% as well as the rest of the prerequisites for REVTeX 4.0
%
% See the REVTeX 4 README file
% It also requires running BibTeX. The commands are as follows:
%
%  1)  latex apssamp.tex
%  2)  bibtex apssamp
%  3)  latex apssamp.tex
%  4)  latex apssamp.tex
%
%\documentclass[twocolumn,showpacs,preprintnumbers,amsmath,amssymb]{revtex4}
%\documentclass[preprint,showpacs,preprintnumbers,amsmath,amssymb]{revtex4}

% Some other (several out of many) possibilities
%\documentclass[preprint,aps]{revtex4}

%\documentclass[twocolumn,aps,draft]{revtex4}
%\documentclass[prb]{revtex4}% Physical Review B

% Include figure files
%\usepackage{dcolumn}% Align table columns on decimal point
%\usepackage{bm}% bold math

%\nofiles

%\newcommand{\bra}[1]{\left\langle {#1} \right|}
%\newcommand{\ket}[1]{\left|  #1 \right\rangle}
%\newcommand{\bracket}[3]{\langle {#1} | {#2} | {#3} \rangle}
%\newcommand{\braket}[2]{\langle {#1} | {#2} \rangle}
%\newcommand{\aver}[1]{\langle {#1} \rangle}
%\newcommand{\abs}[1]{\left| {#1} \right|}
%\newcommand{\raw}[0]{\rightarrow}
\begin{widetext}
	%\begin{document}
	\newpage

	%	\preprint{XXX}
	\begin{center}
		\textbf{\large Supporting Information: Motion Transduction with Thermo-mechanically Squeezed Graphene Resonator Modes}
	\end{center}
	
	%\title{Exciting high-Q graphene oscillations via coupled substrate modes }
	
	\author{Rajan Singh}
	\affiliation{Department of Physics, Indian Institute of Technology - Kanpur, UP-208016, India}
	
	\author{Ryan J.T. Nicholl}
	\affiliation{Department of Physics and Astronomy, Vanderbilt University, Nashville, Tennessee 37235, USA}

	\author{Kirill Bolotin}
	\affiliation{Department of Physics, Freie Universitat Berlin, Arnimallee 14, Berlin 14195, Germany}
	
	\author{Saikat Ghosh}
	\email{gsaikat@iitk.ac.in}
	\affiliation{Department of Physics, Indian Institute of Technology - Kanpur, UP-208016, India}

	\date{\today}

	%\pacs{xxx}
	
	\maketitle
%	\begin{center}
%		\text{*Email: gsaikat@iitk.ac.in}
%	\end{center}

	%	\section{Effective mass of resonators}
	%	Effective mass of circular membranes is mode dependent \cite{Hauer13}. 
	%	For fundamental mode of single layer graphene (SLG) with diameter $2R=15\mu m$, it is
	%	\begin{equation}
	%		m_{g}=0.2695\times\rho\pi R^2 t=0.353\times10^{-16} kg
	%	\end{equation}
	%	Here $\rho$ and $t$ are density and effective thickness of graphene respectively. Similarly, for fundamental mode of $20\mu$m diameter SLG, it is 
	%	\begin{equation}
	%		m_{g}=0.2695\times\rho\pi R^2 t=0.625\times10^{-16} kg
	%	\end{equation}
	%	
	%	In case of  square membrane effective mass is independent of mode.
	%	For a $\rm SiNx$ membrane with dimensions $320 \times 320 \times 0.3 \mu m^3 $ and density $\rho_s=3100 kg/m^3$, we get 
	%	\begin{equation}
	%		m_{s}=\frac{1}{4}\rho_s l w h=2.38\times10^{-11} kg
	%	\end{equation}
	
	\section{Experimental methods:}
	\emph{ Experimental setup:}
	Motion of graphene and ${\rm Si}{\rm Nx}$ mechanical resonators is detected with a confocal microscope (spot-size of $2~{\rm \mu m}$) (Fig. 1b), using a probe laser (ECDL,Toptica) at a wavelength of $780~{\rm nm}$. The microscope forms one arm of a Michelson interferometer while a second (reference) arm is actively stabilized using PI lock box to counter ambient vibrations. The probe, derived from a frequency and power stabilized external cavity diode laser, is detected with a balanced photo-detector of bandwidth of $45~{\rm MHz}$. Subsequently, a spectrum (network) analyser is used to detect (drive) displacement power spectrum of the graphene or ${\rm Si}{\rm Nx}$ target resonator.
	The sample is placed in a rough vacuum chamber (at a pressure of $\sim 10$ mTorr), along with high voltage gate contacts. The chamber in turn is placed on a 3D scanning stage (Thorlabs) having active position locking, with a closed-loop position stability of $5~{\rm nm}$.\\
	
	\noindent
	\emph{Sample Preparation:} Silicon nitride membranes (thickness 300 nm) are fabricated by depositing low-stress silicon-rich silicon nitride on both sides of a silicon chip. An array of holes of 15 and 20 um diameter is then patterned in the nitride using standard fabrication procedures. A metallic contact (20 nm Au) is deposited onto the top surface of the SiNx to facilitate electrical gating. Monolayer chemical vapor deposition (CVD) graphene is then transferred onto holes in the nitride membranes. We use a high-quality atmospheric CVD growth and wet transfer. The samples are subsequently annealed in an $Ar-H_2$ environment at $350^\circ C$. The graphene membranes remained clamped to the sample chip via van der Waals interactions forming suspended circular graphene membranes.	
	\section{Experimental parameters:}
	
	\subsection{Displacement Calibration from Thermal modes of graphene resonator}
	
	A vibrational mode of a suspended graphene resonator, driven thermally, can be modeled as a simple harmonic oscillator \cite{HauerB13}:
	\begin{equation}
		\ddot{x}_g+ \gamma_g\dot{x}_g+ {\omega_g}^2 x_g = F_g^{th}/m_g, 
	\end{equation}
	
	%Assuming the solution to be of form 
	%\begin{equation}
	%x_k\sim \tilde{x}_k\exp(i\omega_gt)
	%\end{equation} 
	
	% or equivalently, for Fourier components of graphene displacement $ \tilde{x}_g $ and force $\tilde{F}_g$.
	%\begin{equation}
	%	(\omega_g^2-\omega^2+i\gamma_g \omega)\tilde{x}_g=\frac{\tilde{F}_g}{m_g}.
	%\end{equation} 
	
	The corresponding thermal displacement power spectrum then takes a form
	\begin{equation}
		S^{\rm th}_{x,g}=\left|\tilde{x}_g(\omega)\right|^2 = \left|\chi_g\right|^2\frac{S^{\rm th}_{F,g}}{m_g^2}
	\end{equation}
	where force spectral density $ S^{\rm th}_{F,g} = 4 k_B T m_g\gamma_g $. Here $\chi_g $ is a linear response function: 
	$\chi_g =  {1}/[{({\omega_g}^2-\omega^2)+i\gamma_g\omega}]$. 
	
	%Accordingly, in terms of natural frequency ($\nu$) and damping, we get
	%\begin{equation}
	%	S_{x,g}= \frac{k_{B}T\nu_{g}}{2\pi^{3}m_{g}Q_{g}[(\nu^{2}-\nu^{2}_{g})^{2}+(\frac{\nu\nu_{n}}{Q_{g}})^{2}]}.
	%\end{equation} 
	
	On the contrary, the measured voltage power spectrum ($S_{v,g}(\omega)$) is expressed as
	\begin{equation}
		S_{v,g}(\omega)=\kappa S^{\rm th}_{x,g} +S_{noise},
	\end{equation}  
	
	with $S_{noise}$ as the noise power introduced by our measurement setup and $\kappa$ (measured in units of $\rm V^2/m^2$) is a voltage to displacement conversion factor.
	By fitting the Eq. S4 to experimentally measured thermal noise spectrum of the graphene drum, we extract the value of $\kappa$, $\omega_{g}$, $ Q_g $(quality factor of graphene), and  $S_{noise}$.
	To extract thermal displacement spectrum (in units of $\rm m/\sqrt{Hz}$), experimental data (units of $\rm V/\sqrt{Hz}$) is divided by  $\sqrt{\kappa}$.\\ 
	For the fundamental mode of a graphene resonator of diameter $20$ $\rm \mu m$, we use: $m_{g} =29\times 0.625\times10^{-16}$ kg for fitting.

	\begin{figure*}\includegraphics[scale=0.7]{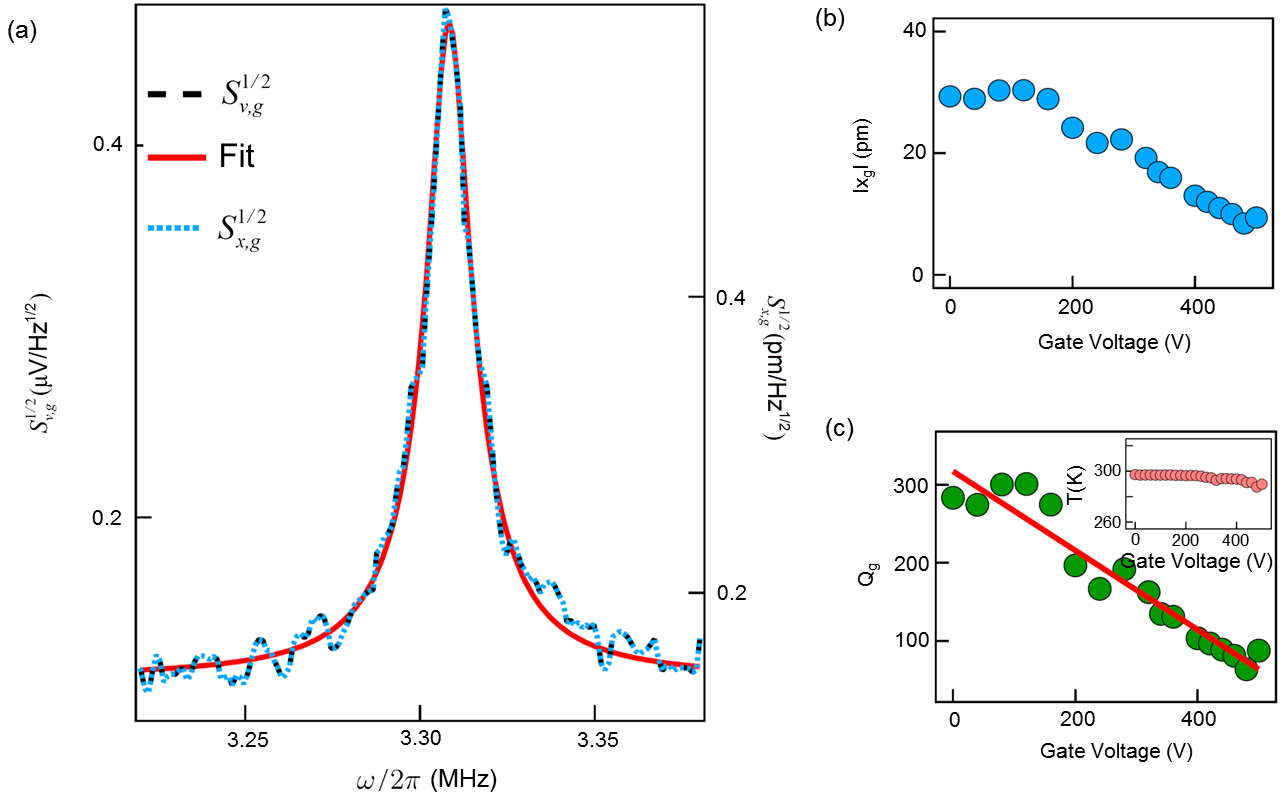}
		\caption{\textbf{ Displacement calibration: (a)} Displacement spectral density (right axis) and voltage spectral density (left axis) vs. frequency for thermally excited graphene drum (blue curve) together with a fit to Eq. S.3 (red curve). \textbf{(b)} Displacement amplitude vs. gate voltage for the same device. \textbf{(c)} Quality factor vs. gate voltage along with a linear fit to the data.\textbf{(inset)} Temprature corresponding to the fundamental mode of graphene with the gate voltage calculated using equipartition theoram }
	\end{figure*}
	
	Fig. S.1a shows a typical of the voltage spectrum along with fitting Lorentzian function and corresponding displacement spectrum. In Fig. S.1b, we plot the extracted thermal displacement ($|x_g|$) of the fundamental mode which decreases with increasing gate voltage, in accordance with equipartition theorem. There is a corresponding monotonic decrease in the extracted quality factors (Fig. S.1c).   
	
	\subsection{Brownian modes of Graphene}
	Fig. S.2a shows a typical Brownian spectrum ( $20$ $\rm \mu m$ resonator, device A) with 8 visible graphene modes. Increasing d.c. gate voltage results in overall blue shift in mode frequency (with small initial red shift for few higher modes)~\cite{Chen-09}.
	Fig. S.2b shows power spectrum at $V_g=100$ V, along with expected values for an ideal circular graphene resonator (blue dashed lines). Interestingly, we observe splitting of few modes. We attribute this to asymmetric tension in graphene and was verified numerically (COMSOL). Simulated resonance modes of graphene with varying the anisotropy ($\Delta T $) shows splitting in modes with axes of symmetry. Mode frequencies for first 10 modes are plotted in Fig. S.2c. The simulation results are in overall agreement with the observations, explaining the emergence of splitting with increasing $\Delta T$. 
	We estimate the mass and tension of the graphene drum by fitting the dispersion of fundamental mode (Fig. S.2d) following Continuum Mechanics model~\cite{Chen-09,Chen-13}. The extracted mass is $29.18(\pm 0.02)m_0$ (where $m_0 = 0.625 \times 10^{-16} \rm kg$ is assumed to be the mass of pristine residue-free graphene) and tension(T) is $3.588(\pm 0.004)\times 10^{-4} \rm N/m$.
	%\section{Anisotropy in the graphene drum}
	
	%\begin{figure*}\includegraphics[scale=.7]{S2}
	%	\caption{\textbf{Brownian motion in Anisotropic Graphene Drum}\textbf{(a)}Tension variation of first five mode of graphene with the gate voltage \textbf{(Inset)}  shows the change in anisotropic tension of $f_{(1,2)} $ and $f_{(3,4)} $ with gate voltage .\textbf{(b)}Theoretical plots of lift in degeneracy in mode with nodal line with  the anisotropic tension }
	%\end{figure*} 
	
	\begin{figure*}\includegraphics[scale=.7]{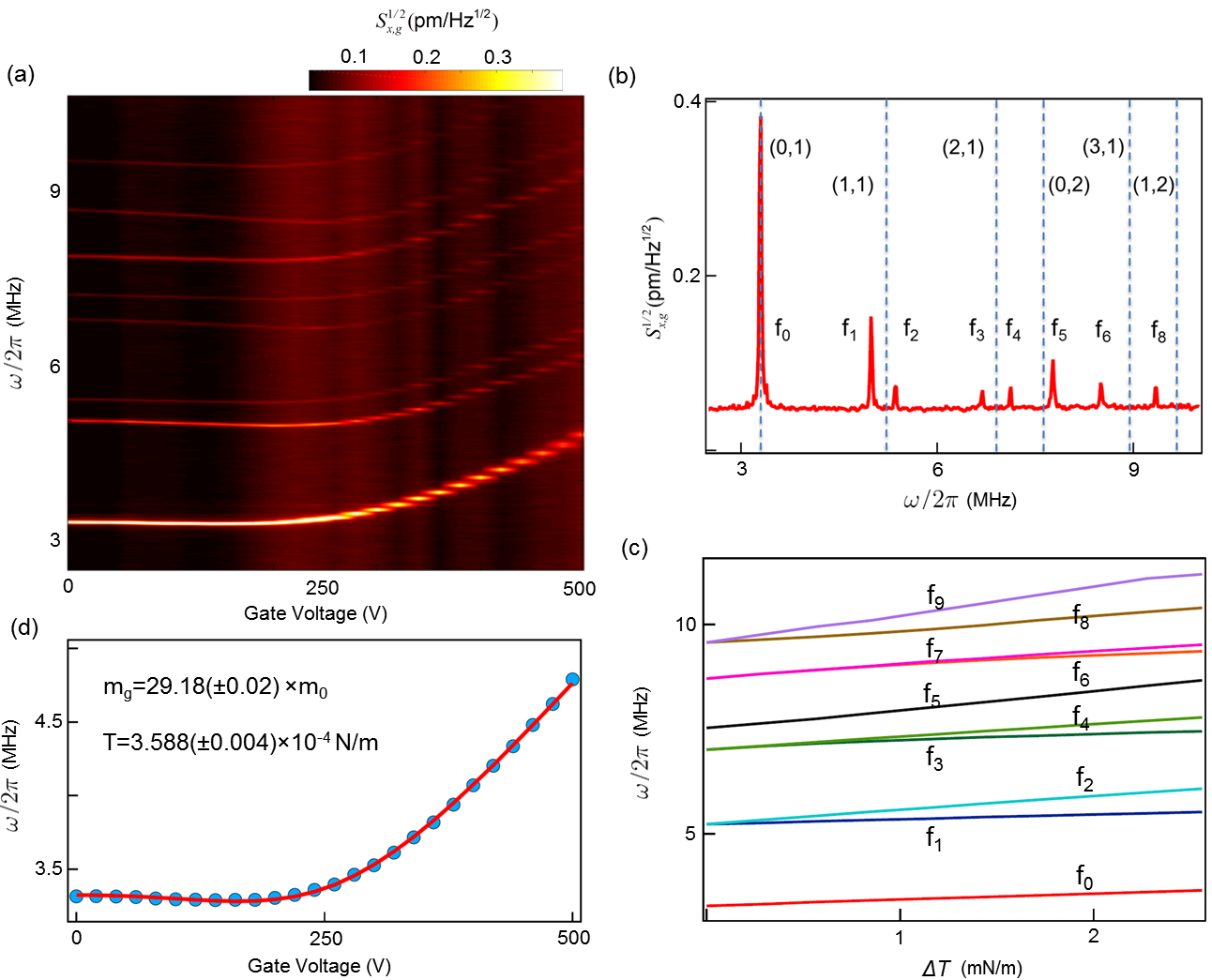}
		\caption{\textbf{Brownian modes with anisotropic tension:} \textbf{(a)} Measured Brownian displacement spectral density with varying gate voltage for a un-driven graphene resonator. \textbf{(b)} A cross-section (for the data of (a)) at $V_g=100$ V, along with expected values for a circular membrane (blue dashed lines). \textbf{(c)} Simulated variation of mode frequencies of a circular graphene resonator vs. tension anisotropy ($\Delta T$). \textbf{(d)} Estimation of mass and tension by fitting the dispersion of fundamental mode of graphene with gate voltage.}
	\end{figure*} 

	\begin{figure*}\includegraphics[scale=0.7]{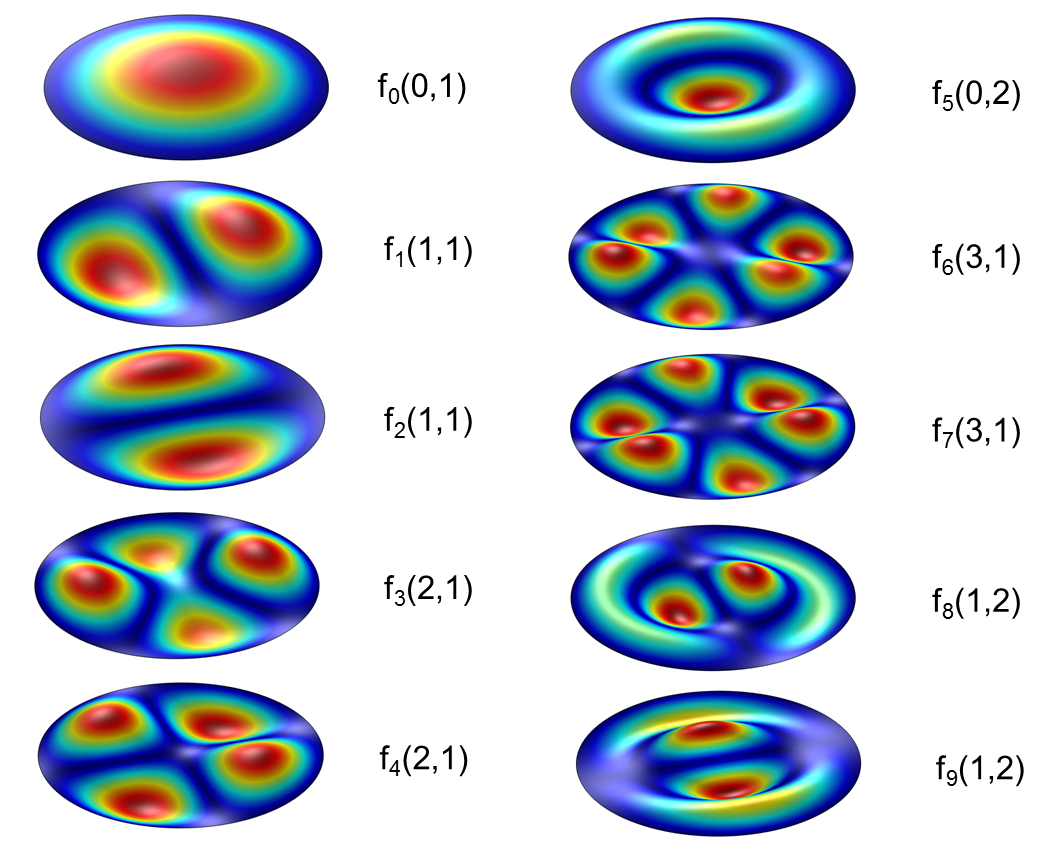}
		\caption{\textbf{Graphene spatial modes:} Simulated spatial mode profiles of a circular resonator under anisotropic tension. $f(m,n)$ denotes the frequency corresponding to the mode index $(m,n)$.}
	\end{figure*}

	\subsection{Silicon Nitride resonator}
	Graphene is supported on holes etched on a large area SiNx plate which has its own distinct vibrational modes. For the $\rm SiNx$ resonator of effective dimension of $320 \times 320 \times 0.3$ $\rm \mu m^3 $ and density $\rho_s=3100$ $\rm kg/m^3$, we estimate its mass $m_{s} = 2.38\times10^{-11}$ kg. These values yield a thermal displacement power spectrum $S_{x,s}^{1/2} = \left( \frac{4k_B T}{ m_s \gamma_s \omega_s^2}\right)^{1/2} \simeq 10 $$\rm$ $\rm fm/\sqrt{Hz} $  (at $\omega=\omega_s$, at a SiNx resonant frequency $\omega_s=2\pi \times 4.163$ MHz with a damping $\gamma_s=2\pi \times1.5$ kHz). We do not observe such Brownian oscillations of the SiNx modes as it lies below our detection noise floor. 
	
	\subsection{Driven SiNx modes }
	
	We excite SiNx resonator modes photo-thermally, using a laser (DL pro, Toptica), which is amplitude modulated with a AOM and is incident at an angle.
	%
	%\begin{figure*}\includegraphics[scale=0.6]{S5}
	%	\caption{\textbf{SiN modes of vibration} \textbf{(a)} compares the driven modes of vibration of $\rm SiN$ at inbuilt tension 600 MPa (Device:1) and 150 MPa (Device:2) \textbf{(b)} and \textbf{(c)} shows  few modes of vibration of Device:1 and device:2 with in range of 4-4.5 MHz respectively}
	%\end{figure*}
	When SiNx is driven with the modulated laser beam, while the probe is focused on the same device, we observe a zoo of modes. These modes remain unaffected with gate voltage (Fig. 1d of the main text), are of relatively high quality factors ($Q_s \sim 10^3$) with eigenfrequencies $f_{mn} = \sqrt{T_s/4\rho_s L^2}\sqrt{m^2+n^2}$ of that of a square membrane. Furthermore, the density of vibrational modes can be controlled by changing its inbuilt tension ($T_s$). In Fig. S.6a, the device with high built-in tension (device A) has smaller mode density as compared to low built-in tension (device B). Fig. S.6b represents hybrid mode resulting from interaction of graphene with many modes of SiNx in device B.

	\begin{figure*}\includegraphics[scale=0.7]{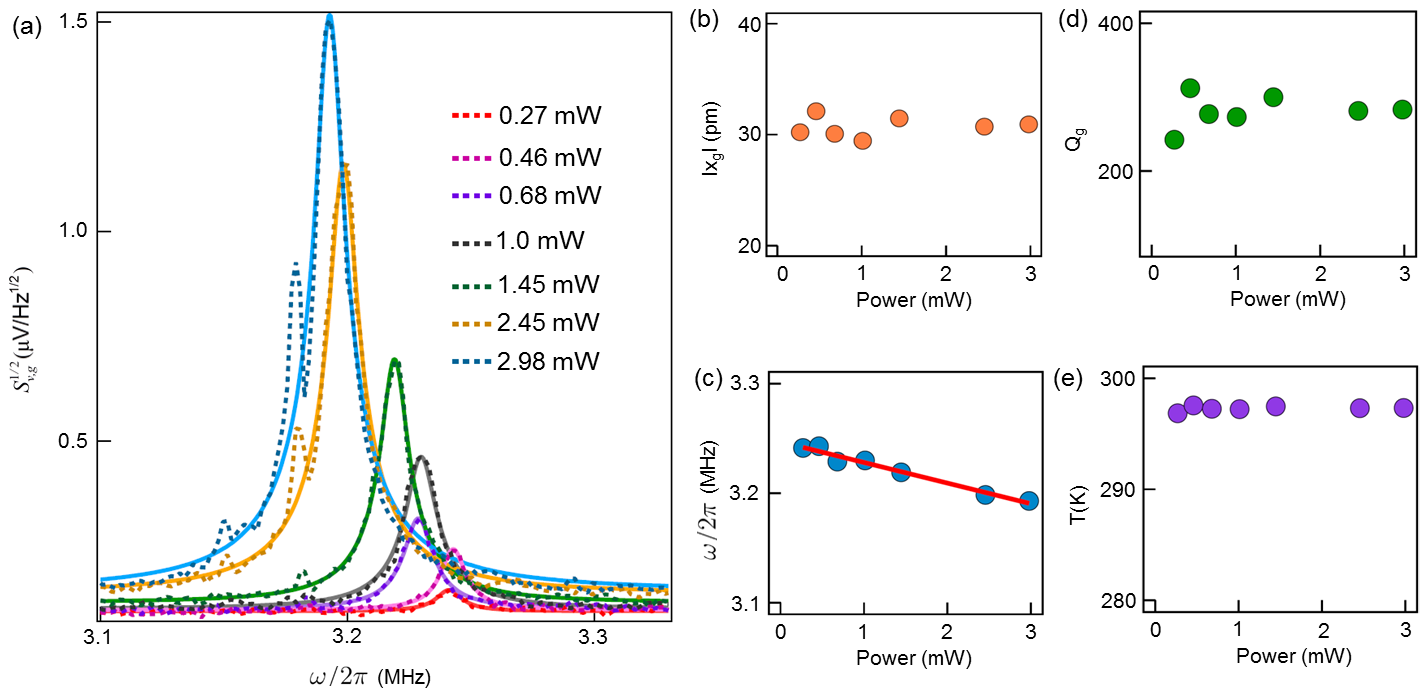}
		\caption{\textbf{Power spectra at different probe powers: (a)} Voltage spectrum of graphene fundamental mode at different probe powers together with fits to the data. Absolute displacement \textbf{(b)}, resonance frequency \textbf{(c)},  quality factor \textbf{(d)}, and temperature \textbf{(e)} extracted from the fit vs. excitation power. Note that displacement, quality factor, and temperature remain unchanged while frequency is redshifted $\sim 50$ $\rm kHz$.  }
		
	\end{figure*}

	\subsection{Graphene spectrum with probe power}
	
	We studied dependence of thermal motion of graphene's fundamental mode with increasing incident probe power that was varied from 0.1 $\rm m$W to 3 mW and observed changes in displacement, frequency, quality factor and temperature by fitting the mode profiles.
	
	We found that the absolute displacement  for the fundamental mode  of the graphene remains constant ($\sim 30$ $ \rm pm$) when the probe power is increased (Fig. S.4b), while the resonance frequency redshifts by $\sim$ 50 kHz (Fig. S.4c). We also estimated the effective mode temperature using equipartition theorem and have not observed significant changes with probe power below 1 mW (Fig. S.4e). Based on these observations, we keep probe power less than 0.5 $\rm mW$ for all subsequent measurements.

	\begin{figure*}\includegraphics[scale=0.7]{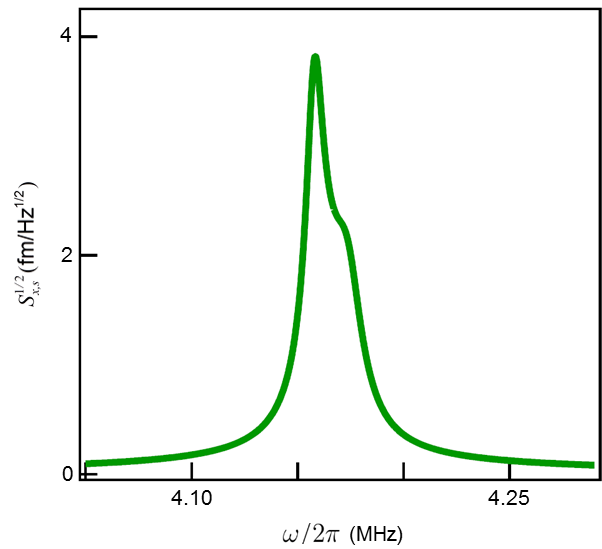}
		\caption{\textbf{Extracted displacement spectral density of hybrized SiNx:}  Extracted SiNx Brownian spectrum, hybridized with graphene as extracted from Eq. S.9 using the fitting parameters from Eq. S.8 and Fig. 2a of main text(probe power at 0.1 mW). }
		
	\end{figure*}
	
	\section{On the theoretical model:}
	
	To describe graphene/SiNx interaction~\cite{SchwarzC16,Okamoto13}, we consider a simple model of bilinear coupling of two harmonic oscillators with an effective interaction Hamiltonian $\hat{H}_{int}=\alpha \hat{x}_g\hat{x}_s$. The corresponding coupled equations are:
	\begin{subequations}\label{first: a}
		
		\begin{equation}\label{first: a}
			\ddot{x}_g+ \gamma_g\dot{x}_g+ {\omega_g}^2 x_g - \frac{\alpha}{m_g} x_s= \frac{F_{g}}{m_g} 
		\end{equation}
		and	
		\begin{equation}\label{first: b}
			\ddot{x}_s + \gamma_s\dot{x}_s+ {\omega_s}^2 x_s - \frac{\alpha}{m_s} x_g=\frac{F_{s}}{m_s},   
		\end{equation}
	\end{subequations}
	where $F_{g}$ and $F_{s}$ are forces acting on graphene and $\rm SiNx$ oscillators respectively.
	
	Solving the above coupled equations, we get
	\begin{equation}
		S_{x,g}=	\left|\tilde{x}_g(\omega)\right|^2 = \left|\frac{\chi_g}{1-\frac{\alpha^2}{m_g m_s}\chi_g\chi_s}\right|^2\frac{\langle F_{g}^2 \rangle}{{m_g}^2} + \frac{\alpha^2}{m_g^2}\left|\frac{\chi_g\chi_s}{1-\frac{\alpha^2}{m_g m_s}\chi_g\chi_s}\right|^2 \frac{\langle F_{s}^2 \rangle}{{m_s}^2}
	\end{equation}
	\begin{equation}
		S_{x,s}=	\left|\tilde{x}_s(\omega)\right|^2 = \left|\frac{\chi_s}{1-\frac{\alpha^2}{m_g m_s}\chi_g\chi_s}\right|^2\frac{\langle F_{s}^2 \rangle}{{m_s}^2} + \frac{\alpha^2}{m_s^2}\left|\frac{\chi_g\chi_s}{1-\frac{\alpha^2}{m_g m_s}\chi_g\chi_s}\right|^2 \frac{\langle F_{g}^2 \rangle}{{m_g}^2}
	\end{equation}
	
	Here $\chi_{g,s}$ are the response functions (susceptibilities, scaled with the respective masses) of the bare graphene and SiNx oscillators, respectively and are defined as
	\begin{equation}
		\chi_{k} =  \frac{1}{({\omega_k}^2-\omega^2)+i\gamma_k\omega},
	\end{equation}  
	where $k=g,s$. Rewriting, we get
	\begin{equation}
		S_{x,g}=\frac{C[(\omega_{s}^2-\omega^2)^2+\gamma_{s}^2\omega^2]+D[\frac{\alpha^2}{m_g^2}]}{[(\omega_{g}^2-\omega^2)(\omega_{s}^2-\omega^2)-\gamma_{g}\gamma_{s}\omega^2-\frac{\alpha^2}{m_g m_s}]^{2}+[(\omega_{g}^2-\omega^2)\gamma_{s}\omega+(\omega_{s}^2-\omega^2)\gamma_{g}\omega]^{2}},
	\end{equation} 
	
	\begin{equation}
		S_{x,s}=\frac{D[(\omega_{g}^2-\omega^2)^2+\gamma_{g}^2\omega^2]+C[\frac{\alpha^2}{m_s^2}]}{[(\omega_{s}^2-\omega^2)(\omega_{g}^2-\omega^2)-\gamma_{s}\gamma_{g}\omega^2-\frac{\alpha^2}{m_g m_s}]^{2}+[(\omega_{s}^2-\omega^2)\gamma_{g}\omega+(\omega_{g}^2-\omega^2)\gamma_{s}\omega]^{2}},
	\end{equation} 
	
	where C $= \frac{\langle F_{g}^2 \rangle}{m_{g}^2}$ and D $= \frac{\langle F_{s}^2 \rangle}{m_{s}^2}$.
	%\begin{equation}
	%S_{xx}^{1/2}=\sqrt{\frac{C[(w_{s}^2-w^2)^2+\gamma_{s}^2w^2]}{[(w_{g}^2-w^2)(w_{s}^2-w^2)-\gamma_{g}\gamma_{s}w^2-\frac{\alpha^2}%{m_g m_s}]^{2}+[(w_{g}^2-w^2)\gamma_{s}w+(w_{s}^2-w^2)\gamma_{g}w]^{2}}}
	%\end{equation} 
	
	Equation S.8 fits well to the power spectra of thermally excited graphene-substrate coupled mode (Fig. 2a of main text). The fitting coefficeents are:
	$ C = 103.31 (\pm 5.61)\times 10^{-5}$ $ \rm N^2/Kg^2$, $D = 73.97(\pm 9.78)\times 10^{-10}$ $\rm N^2/Kg^2$, $\omega_{g}/2\pi =4.1688 (\pm 0.0003)$ $\rm \rm MHz$,$\omega_{s}/2\pi =4.1633 (\pm 0.0003)$ $\rm MHz$, $\gamma_g/2\pi =0.021445 (\pm  0.000833)$ $\rm MHz$, $ \gamma_s/2\pi=0.001502(\pm0.000508)$ $\rm MHz$, $\alpha/4\pi^2=1.1919 (\pm0.0023)\times10^{-14}$ $\rm Kg MHz^2$. Figure S.5 shows the corresponding extracted SiNx spectrum, as extracted from Eq. S.9, using these parameters.
	\\

	\begin{figure*}\includegraphics[scale=0.7]{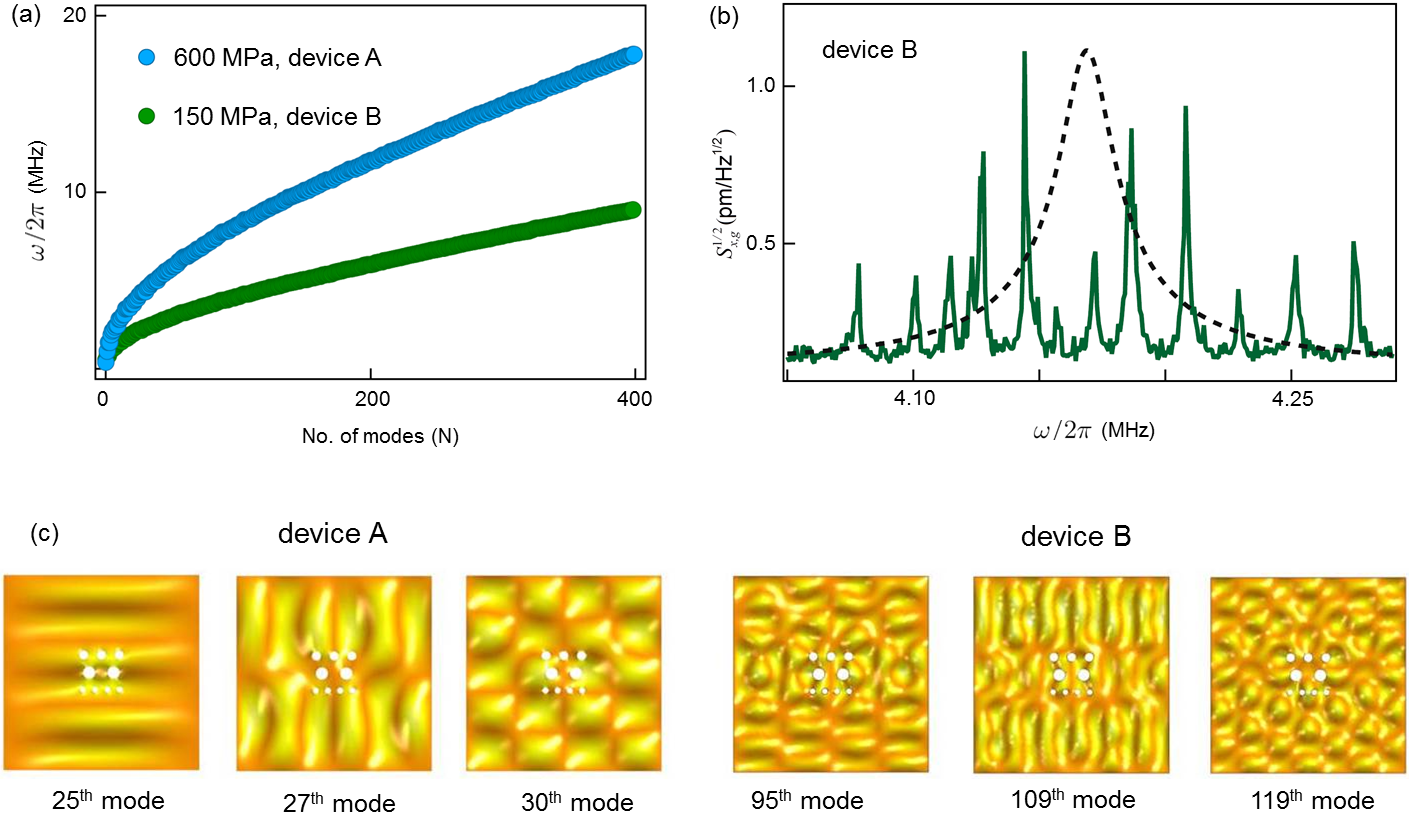}
		\caption{\textbf{SiNx modes:} \textbf{(a)}  COMSOL simulation of SiN modes frequency at built-in tension corresponding to device A (600 MPa) and device B (150 MPa). \textbf{(b)} Interaction between a graphene mode and SiN modes for a low built-in tension device 2. \textbf{(c)} COMSOL simulation of spatial profiles of SiN resonance modes.} 
	\end{figure*}

	\section{Motion transduction:}
	
	The SiNx target oscillator is coupled to a thermal bath, which sets its decay rate $\gamma_s$ along with a fluctuating thermal force $F_s^{th}$. Additionally, it is driven externally with a phase-coherent drive $F_d$ such that the total force acting on the oscillator takes a form $F_s(t)=F_s^{th}+F_d(t)$.
	
	The equation of motion of the uncoupled SiNx target oscillator is 
	\begin{equation}
		\ddot{x}_s+ \gamma_s\dot{x}_s+ {\omega_s}^2 x_s = F_s/m_s 
	\end{equation}
	\subsection{Interferometric detection scheme}
	When this target oscillator is directly probed with an interferometer, the intensity $I(t)$ at the detector takes a form 
	\begin{equation}
		I(t)=I_s(t)+I_r(t)+2\sqrt{I_sI_r}sin(\phi_s+\phi_r).
	\end{equation}
	
	Here $\phi_s (=\frac{2\pi}{\lambda}x_s(t)$) and $\phi_r$ are phases corresponding to signal and  reference arms with intensities $I_s(t)$ and $I_r(t)$, respectively. For small target oscillator displacements
	compared to wavelength, $x_s(t)\ll \lambda$, we approximate $\sin\phi(t)\sim\phi(t)$,  such that
	
	%\begin{equation}
	%	I(t)=I_s(t)+I_r(t)+2\sqrt{I_sI_r}[\frac{2\pi}{\lambda}x_s cos(\phi_r)+sin(\phi_r)]
	%\end{equation}
	\begin{equation}
		I(t)=I_s(t)+I'_r(t)+\zeta_s x_s
	\end{equation} 
	
	where $\zeta_s=\frac{4\pi}{\lambda}\sqrt{I_sI_r}\cos\phi_r$ is the interferometric gain coefficient  and $I'_r(t)=I_r(t)+2\sqrt{I_sI_r}sin\phi_r$, is the modified reference intensity. The phase of the reference part is denoted as the lock phase. By changing that phase through control of our active lock circuit (Fig. 1b of the main text), we can optimize the signal to noise ratio.
	
	Power spectrum of the signal recorded by the spectrum analyzer is given as
	\begin{eqnarray}
		S(\omega) &=& \int\langle I(t)I(t+\tau) \rangle e^{-i\omega t} d\tau \\  \nonumber
		&=&  S_{I_0}+\zeta_s^2 S_{x,s}(\omega)
	\end{eqnarray} 
	where $S_{I_0}=S_{I,s}(\omega)+S_{I',r}(\omega)$ is the total spectral noise floor corresponding to optical and technical contributions.

	\subsection{Our proposed detection scheme: two step process}
	
	The amplification scheme proposed here has two parts:
	
	(a) a graphene oscillator mode that couples to the target SiNx resonator mode, thereby amplifying the displacement power spectrum, followed by (b) a Michelson interferometer that detects the amplified graphene displacement.
	
	The corresponding power spectrum recorded at the output of the interferometer then takes a form:
	\begin{equation}
		S_{I}=S_{I_0}(\omega)+\zeta_g^2 S_{x,g}(\omega)
	\end{equation}
	Using the bi-linear model, graphene power spectrum takes a modified form
	\begin{equation}
		S_{x,g}=\frac{|\chi_g|^2}{m_g^2}(\langle F_g^2 \rangle+\alpha^2 \langle x_s(\omega)^2\rangle)
	\end{equation}
	which yields,
	\begin{equation}
		S_{I}=S_{I_0}(\omega)+ \zeta_g^2 S_{x,g}^{th} +\zeta_g^2 \frac{|\chi_g|^2}{m_g^2}\alpha^2S_{x,s}^{th}(\omega)+\zeta_g^2 \frac{|\chi_g|^2}{m_g^2}\alpha^2S_{x,s}(\omega).
	\end{equation}
	Here first three terms are the detection noise, graphene thermal noise and SiNx thermal noise, respectively. These terms determine the overall sensitivity of the graphene/SiNx hybrid. The last term is the target oscillator spectrum multiplied by a coupling and mass dependent pre-factor leading to gain.
	
	The corresponding gain in displacement power spectrum is defined as ratio of displacement spectrum on graphene to that on SiNx, and can be approximated to 
	\begin{eqnarray}
		G_{c}=\frac{\zeta_g^2\left|\frac{\chi_g(\omega)}{m_g} \right|^2 \alpha^2 S_{x.s}(\omega)}{\zeta_s^2  S_{x,s}(\omega)}
		\simeq \left|\frac{\alpha}{m_g\gamma_g\omega_g}\right|^2.
	\end{eqnarray} 	
	\subsubsection{Gain in device A}
	The reflectivity ($ R=I_{g,s}/I_i $, where $I_{g,s}$ is the reflected intensity from graphene or SiNx, and $I_i$ is the incident intensity of probe beam) of SiNx is 15.2 times larger than that of graphene. To compensate for the difference in reflectivity we divide the SiNx power spectrum  by 15.2. In other words, detected intensity of SiNx is 15.2 times larger compare to that from graphene.  
	Comparing graphene spectrum, $S_{x,g}(\omega)$ with that of bare ${\rm Si}{\rm Nx}$, $S_{x,s}(\omega)$  we estimate the average gain to be $G_c \sim 6.21\times 10^3$ ($38~{\rm dB}$) (Fig. 2b, main text). 
	\subsubsection{Gain in device B}
	As discussed earlier, SiNx in device B has low built-in tension  resulting in higher mode density leading to interaction of a single graphene mode with multiple SiNx modes.
	Fig. S.7 shows the gain corresponding to three modes in device B while inset shows corresponding weakly driven SiNx mode. Coupling to graphene results in average gain of 36.4 dB. We ascribe variations to varying coupling strengths between SiNx modes and graphene. 
	
	\subsection{Backaction on SiNx}
	Corresponding power spectrum of the SiNx target oscillator takes a form:
	\begin{equation}
		S_{x,s}(\omega)=\left|\frac{\chi_s}{m_s}\right|^2[F_s+\alpha x_g(\omega)]^2	=\left|\frac{\chi_s}{m_s}\right|^2S_{F_s}+\left|\frac{\chi_s}{m_s}\right|^2\alpha^2 S_{x,g}(\omega),
	\end{equation}
	where the first term in r.h.s is due to the external drive ($F_s$) acting on SiNx. The second term is due to graphene's \textit{backaction} on SiNx, such that $S_{x}^{ba} =\left|\frac{\chi_s}{m_s}\right|^2\alpha^2 S_{x,g}(\omega)$.
	
	%For external (photothermal) drive (Fig. 2b of the main text), we estimate  ${S_{x}^{ba}}^{1/2}=207 (\pm 40)$ $\rm fm/\sqrt{Hz}$ for device A, on resonance.\\
	Backaction corresponding to thermal fluctuations of graphene ($S^{\rm th}_{x,g}$) leads to
	${S_{x}^{ba}}^{1/2} =\frac{\alpha}{m_s\gamma_s\omega_s} {S_{x,g}^{th}}^{1/2} \sim 31 (\pm 5) $ $\rm fm/\sqrt{Hz}$.

	\begin{figure*}\includegraphics[scale=0.7]{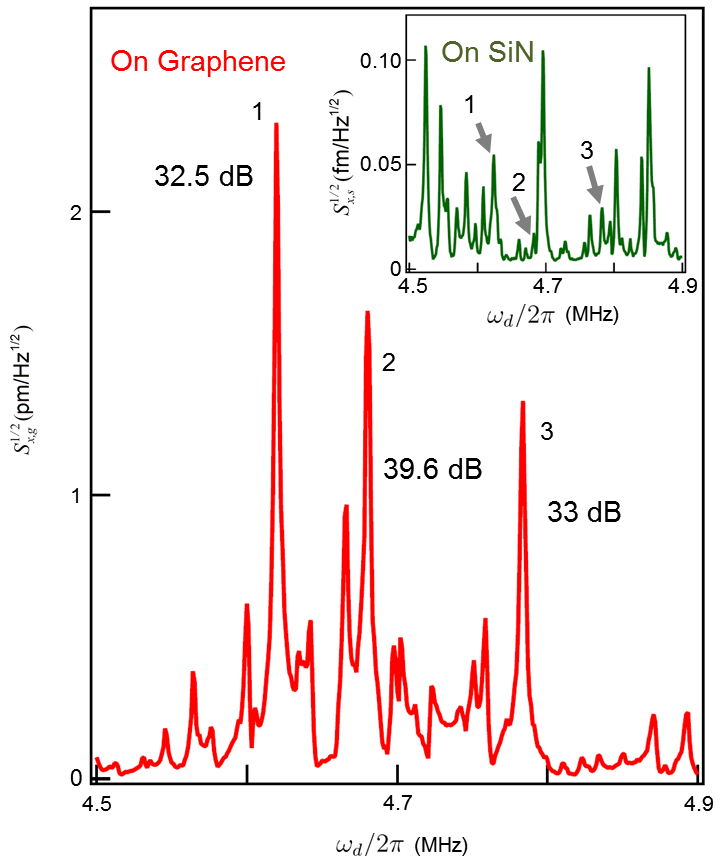}
		\caption{\textbf{Gain in device B:} Displacement power spectrum on graphene due to coupling with the SiNx modes (Inset) resulting in massive gain.}

	\end{figure*}
	
	\subsection{Detection bandwidth of graphene} 
	Detection bandwidth of graphene amplifier in device A is $\sim 38$ kHz (FWHM) (Fig. 2a of the main text) whereas in device B due to interaction with more SiNx mode the bandwidth is larger $\sim 196 $ kHz (Fig. S.6b).
	
	\begin{figure*}\includegraphics[scale=0.7]{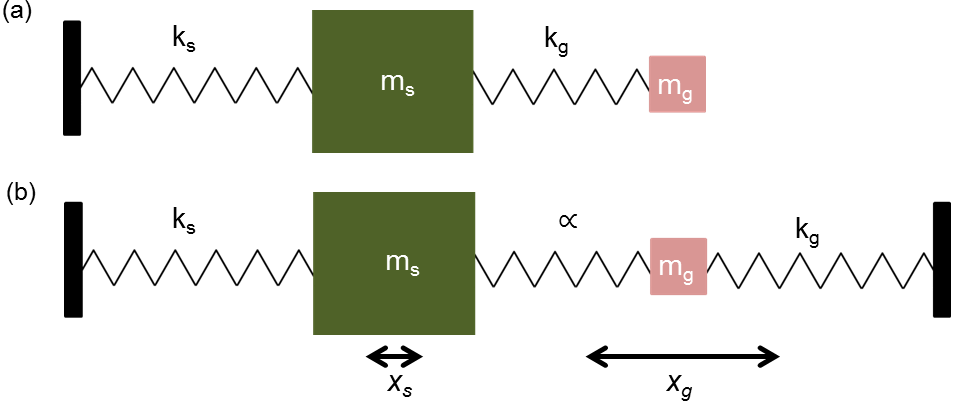}
		\caption{\textbf{Coupled resonator model: (a)} Base excitation model \textbf{(b)},Coupled oscillators with a coupling constant that depends on 2-d geometry.  }
		
	\end{figure*}
	
	\subsection{Validity of our model } 
	
	%The concern raised by Reviewer 1 (R1) is an issue that forms the crux of interpreting the observed data. At the onset, we would like to point out that the gain quoted in the work is defined as the ratio of calibrated displacement power spectrum. Accordingly, as pointed out by R1, for such a scenario, a “base excitation” model will predict a gain of $G = Q^2  \sim 4 \times 10^4 $ (for graphene resonator $Q\sim200$). Our measured Q $ \sim 6.2 \times 10^3$ is lower than that. There is as such no surprise with the large gain: it is lower than the maximum possible gain, which is achieved for coupled oscillators in a base excitation scenario, as pointed out by R1.
	
	%The concern can therefore be rephrased as that regarding the choice of model and interpretation of the data.
	
	The model proposed in the manuscript considers both SiNx and graphene membrane resonators as harmonic
	oscillators, with their single degree of freedom, transverse displacements $x_s$ and $x_g$ respectively,
	characterizing the model. Furthermore, the co-ordinates are assumed to be coupled with a coupling constant
	$\alpha$ and is described with a coupling Hamiltonian $H_{int}=\alpha x_g x_s$ .  The corresponding gain then
	takes a form $G_c \simeq |\frac{\alpha}{m_g \omega_g \gamma_g }|^2$.
	
	One can optimally model the coupled oscillator system as that of graphene oscillator with base excitation of SiNx, to which it is coupled elastically. Such a direct coupling then translates to $\alpha =k=m\omega^2$ and the corresponding gain is then: 
	$G_c \simeq |\frac{\alpha}{m_g \omega_g \gamma_g}|^2=|\frac{\omega_g}{\gamma_g}|^2=Q_g^2 \sim 4\times 10^4$ which would be an order of magnitude larger than the measured $G_c=6.2 \times 10^3$. The differences in  base excitation model~\cite{SchwarzC16,Majorana97} and our  model~\cite{Okamoto13} are particularly evident in a corresponding spring mass picture of Fig. S.8. While the base excitation model considers a direct base excitation (Fig. S.8a), we consider an intermediate coupling (Fig. S.8b), such that the SiNx effective mass spends part of its energy in squeezing a spring with a larger spring constant. The base excitation scenario, when  $\alpha$ reduces to $\alpha=k$ is achieved only in an ideal 1-d scenario, when there is a perfect coupling between the two transverse mode excitations.
	%%
	%
	%Fig. 1(a) Base excitation model. (b) Coupled oscillators with a coupling constant that depends on 2-d geometry.
	%
	However, for 2-d membranes (or plates with a finite thickness), along with coupling to transverse mode, base excitation due to clamped graphene drums, results in longitudinal and radial excitation modes. Such coupling are particularly critical due to the following reasons:
	\begin{enumerate}
		\item Both SiNx and graphene resonators have large intrinsic and anisotropic tension.  Effective coupling of transverse modes of such plates and membranes have been studied extensively in connection to MEMS and NEMS and is found to scale down with increasing intrinsic tension.
		\item Since the vibrational mode of SiNx is high order, the corresponding spatial mode is not uniform, particularly around the edges of graphene. The effective coupling, which depends on the spatial mode overlap of SiNx and graphene at the interface, is thereby non-optimal.
		\item Folds on graphene leading to hidden areas and accordingly, more complex mode structures than simple circular membrane.
		\item A part of the base excitation energy is spent due to bending rigidity of both SiNx and graphene (albeit small) plates.
	\end{enumerate}

	Such mechanisms, all combined, results in a partial transfer of the transverse mode energy of SiNx to the corresponding transverse mode of graphene. Accordingly, to capture this, we use the model of Fig. S.8b, with an intermediate coupling spring $\alpha \geq k_g$. The larger spring constant physically implies simply that some part of the base excitation is lost to other modes of circular graphene sheet.
	It can also be argued that in such scenarios (with $\alpha$ being a geometric factor for circular plates), the gain $G_c \simeq |\frac{\alpha}{m_g \omega_g \gamma_g }|^2$ indeed scales inversely with the resonator mass, thereby having an advantage in using graphene sheets as transducers.
	Similar to our approach have been used extensively used for two coupled mechanical modes in MEMS and NEMS community before.~\cite{Okamoto13} 
	
	%\section{More on graphene-substrate thermal modes interaction.}
	
	%\subsection{Frequency splitting width with the size of the drum}
	%Splitting width or the coupling strength decreases with the size of the drum.

	%\subsection{Frequency splitting width with the in build tension  }
	%Coupling strength or the frequency splitting width increases with the tension of the drum.

	%\begin{figure*}\includegraphics[scale=0.7]{}
	%	\caption{\textbf{Theoritical plot of frequency of first 10  modes of vibration with the anisotropic tension} }
	%\end{figure*} 

	%\begin{figure*}\includegraphics[scale=0.7]{freq_stress_crop}
	%	\caption{\textbf{Eigen modes of Silicon Nitride membrane  for different value of inbuilt stress}}
	%\end{figure*}
	\begin{figure*}\includegraphics[scale=0.7]{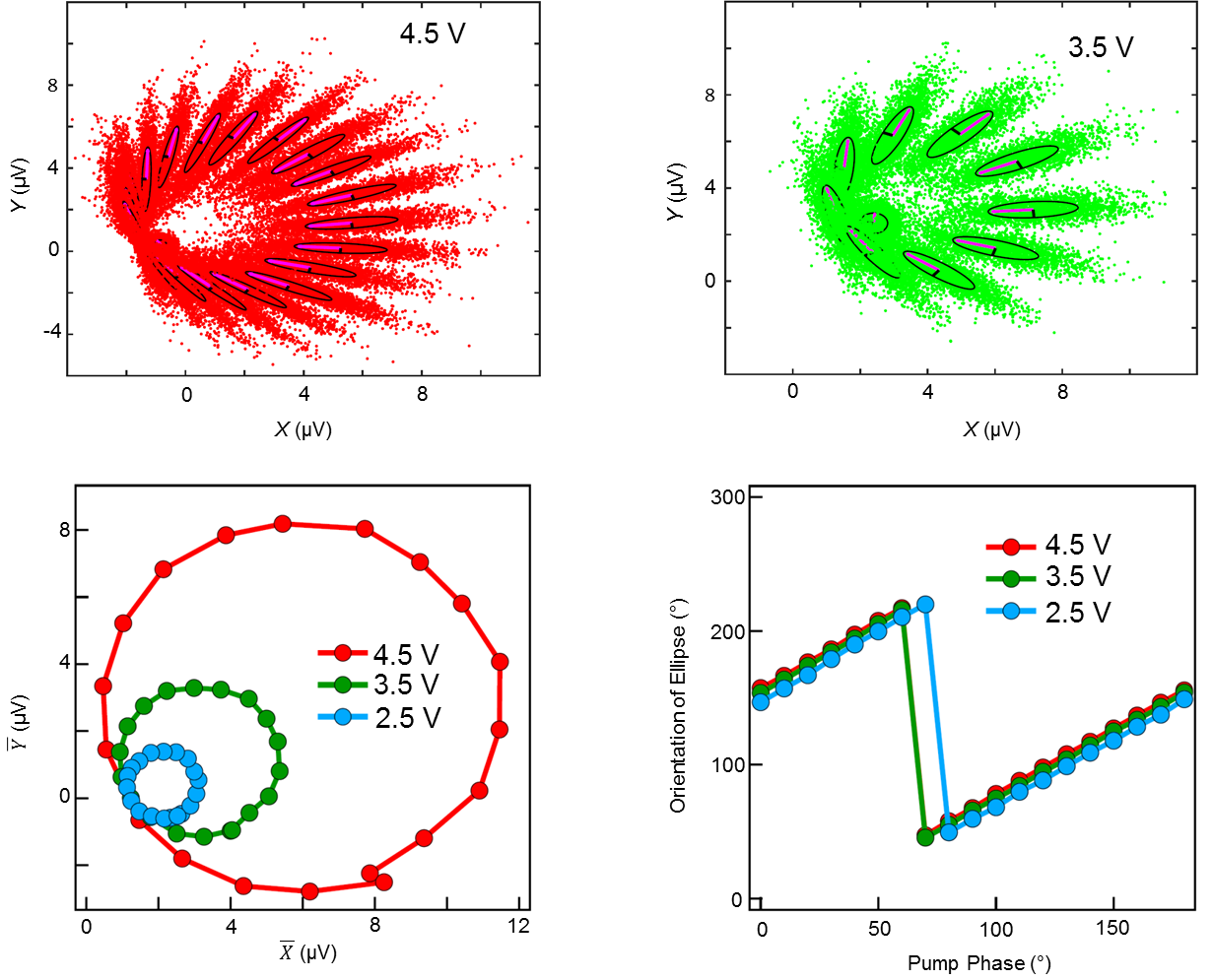}
		\caption{\textbf{Parametic Amplification and Squeezing with pump phase:} \textbf{(a)} Quadrature plots at $\phi_p=0^\circ$ to $180^\circ$ in steps of $10^\circ$ at $V_p^0=4.5$ V  and \textbf{(b)} in steps of $20^\circ$  at $3.5$ V.\textbf{(b)} shows plot of mean of quadtaures $\bar{X}$ and $\bar{Y}$ with $\phi_p$ at 4.5 V, 3.5 V and 2.5 V.\textbf{(c)} Orientation of ellipse with pump phase at different pump voltages.} 
	\end{figure*} 
	\section{Thermo-mechanical Squeezing in Graphene}
	
	The parametric amplification~\cite{RugarD91,Lifshitz08,EichlerN11} can be represented as follows
	\begin{equation}
		m_g\ddot{x}_g+ m_g\gamma_g\dot{x}_g+[k_{g} +k_p(t)]x_g +\eta x_g^2 \dot{x}_g +\beta x^{3}_{g} = F_{g}^{th}+F_{g} cos(\omega_{d}t+\phi_d)
	\end{equation}
	
	where $k$ is the intrinsic spring constant, $k_p(t)(=k_{p}^0cos(\omega_{p}t+\phi_p))$ is the modulated spring constant, $\beta$ is cubic nonlinear coefficient and $\eta$ is nonlinear damping coefficient. In case of degenerate parametric amplification, graphene mode tension is capacitively modulated at twice the resonance frequency ($\omega_{p}=2\omega_{d}) $.
	From above equation, when nonlinear terms are negligible parametric gain is represented as
	\begin{equation}  
		G_{p}=\frac{cos^2(\phi_p+\pi/4)}{(1+\frac{V_p^0}{V_p^c})^2}+\frac{sin^2(\phi_p+\pi/4)}{(1-\frac{V_p^0}{V_p^c})^2}	
	\end{equation}
	which explains the behavior of gain with phase and pump voltage below the critical voltage ($V_p^c$). At $V_p^c$ graphene mode goes into self oscillation regime~\cite{Lifshitz08,EichlerN11}.

	\begin{figure*}\includegraphics[scale=0.7]{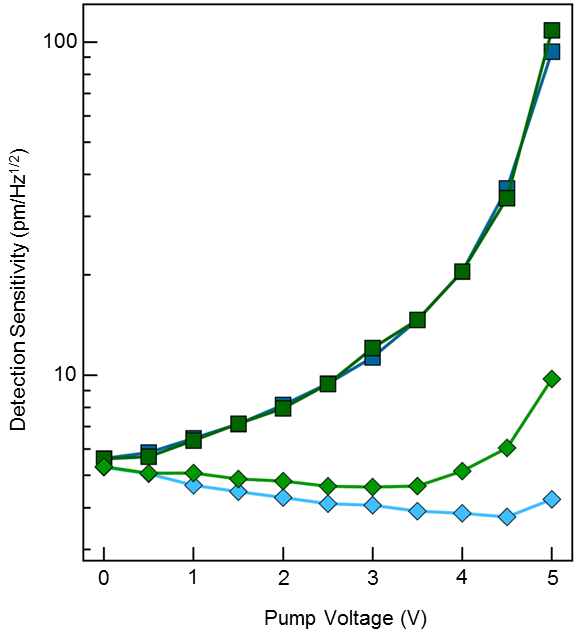}
		\caption{\textbf{Detection sensitivity:} Detection sensitivity with pump voltage along two quadrature at $\phi_p=45^\circ$ (green pair) and $135^\circ$ (blue pair) respectively. We achieved best sensitivity of 3.8 $\rm fm/\sqrt{Hz}$ at $\phi_p=135^\circ$ and $V_p^0$= 4.5 V.} 
	\end{figure*} 
	\subsection{Detection Sensitivity of Graphene after Thermomechanical squeezing}
	
	Detection sensitivity of graphene improves after thermomechanical squeezing along one quadrature at the cost of another.
	Including the contribution from squeezing, the detection sensitivity is written as, $S_{x,min}^{1/2}=\sqrt{\frac{S^{th}_{x,g}\sigma_{X_1}^2+S_{I_0}}{G_{c}}}$
	where $\sigma_x^2$ is the variance or squeezing factor along one axis. In Fig. S.10, detection sensitivity is plotted along both quadratures with pump voltage at $\phi_p=45^\circ$ (green markers) and 135$^\circ$ (blue markers). The best sensitivity we achieved is $3.8$ $\rm fm/\sqrt{Hz}$ at $\phi_p=135^\circ $ and $ V_p^0=4.5$ V where $\sigma_{X_1}=0.58$.

	%\section{Scaling of noises with layers of graphene}
	%$S_x^{imp}=(\frac{4kT\gamma_g}{(2\pi)^3\alpha^2})n_g m_g=1.03\times 10^{-29}n_g$\\
	%$S_x^{ba}=\frac{4kT \alpha^2}{(2\pi)^3m_s^2 \omega_s^2 \omega_g \gamma_s^2 \gamma_g}\frac{1}{n_g m_g}=\frac{1.3\times 10^{-27}}{n_g}$

\end{widetext}

\end{document}